# RECOMMENDATIONS OF THE VAO-SCIENCE COUNCIL
## FOLLOWING THE MEETING OF MARCH 26-27, 2010

THE VAO-SC:
G. FABBIANO (*SAO*, CHAIR), D. CALZETTI (*U. Mass Amherst*), C. CARILLI (*NRAO*), G. DJORGOVSKI (*CalTech*), P. ESKRIDGE (*Minnesota State U.*), Z. IVEZIC (*U. Washington*), E. FEIGELSON (*Penn State U.*), A. GOODMAN (*Harvard*), B. MADORE (*Carnegie*), M. POSTMAN (*STScI*), S. SEAGER (*MIT*), A. SODERBERG (*Harvard*), T. RECTOR(*U. of Alaska*)

## EXECUTIVE SUMMARY

There are already hundreds of terabytes of astronomical data in the public domain, and soon the astronomical community will generate tens of petabytes of new data. There is an enormous science potential to be unleashed by the ability to find relevant datasets, access them, analyze them quantitatively, compare them with theoretical models, and ultimately understand their astrophysical implications. The VAO will enable researchers to (1) analyze data obtained with different instruments over the entire observable spectrum; (2) connect large data sets with high performance computational resources to aid analysis, detect, quantify and visualize hidden patterns, and (3) gain understanding of astrophysical phenomena with the aid of state-of-the art models.

The VAO-SC strongly supports these goals. However, we acknowledge that substantial work is needed to achieve them fully. In particular, the VAO needs to move fast from the 'infrastructure building emphasis' of the NVO phase to be a fully science-enabling organization. To this end:

1. We strongly endorse and support the refocusing of VAO activities on ease of use and science productivity. This focus should include paying immediate attention to user-interface design, tool and data discoverability, interoperability among tools, and visualization strategies.
2. We believe that it is essential to keep an eye toward new tools being developed in the world beyond VAO (IVOA tools, statistical packages, commercial tools, independent tools). We endorse developments to provide the "glue" to incorporate these as rapidly as possible.
3. We support the VAO developing stronger collaborative ties with ground-based and space-based mission, thematic and data centers. And we strongly suggest that the VAO actively engage with the publishers of the astronomical literature in making their data and information VO compatible and generally more easily accessible.
4. We endorse the creation and funding of "VAO Associates" who can collaboratively develop new data sets (e.g. key projects) and tools (e.g. software like astrometry.net) that the VAO community deems scientifically useful judged through a lightweight, rapid, science-driven proposal process.

5. We suggest that the VAO create "friend of the telescope" positions at a few major institutions in the US. Charge these "friends" with supporting the VAO community by:
    a. Helping to establish data centers or "small" data sets not hosted by missions;
    b. Hosting in-person events to discuss user ideas and concerns, and to educate users about new tools;
    c. Creating and maintaining online resources (YouTube-style screen-casts, Wikipedia-style descriptions) demonstrating how to carry out various research functions within the VAO.
6. We acknowledge that the VAO should strive to enable science with VERY large data sets, and in doing so, keep a group focused on data-mining those data sets (in the time, wavelength, and space domains). This activity must be included in the longer-term VAO planning, with well-defined yearly deliverables; it must also be integrated with the five activities listed above.

The VAO-SC thanks the VAO team and the VAO director, Dr. Hanisch, for their presentations and introduction to the VAO, and for suggesting several options of further development, which are the basis for our discussions and recommendations. We thank the funding agencies, NSF and NASA for their support of this project. We are happy for the opportunity to represent the astronomical community in providing science user input to the VAO, and are available to support the planning and development of the VAO with our scientific and technical expertise.

Below are the detailed VAO-SC recommendations, following the March 26-27 meeting at GSFC. Section 1., summarizes the major recommendations of the VAO-SC; Section 2., gives the recommendations for near-term improvements (6 mo. – 1 yr. time scale) needed to ensure uptake of the VAO by the science community; Section 3., addresses science input into the VAO planning process.

# 1. GENERAL RECOMMENDATIONS

**1.1 A Change of Emphasis: from Infrastructure to Science Tools**

The NVO has made admirable progress in `laying the pipes' for convenient user access to complex distributed astronomical databases. However, early use of the system for top-quality science has been weak, pointing to a lack of value-added intermediate software products that assist the user in extracting knowledge and understanding from the extracted data. Existing IVOA science analysis capabilities (such as TopCat, VOPlot and VOStat) have variable quality, and are not well integrated into the NVO software environment. While realizing that some infrastructure work will be necessary in the VAO, the move from the infrastructure-building NVO to the operational VAO must emphasize the transition to science enabling capabilities. In particular,
- We strongly endorse the proposal by the VAO team to provide soon important capabilities for the user community. In particular, providing soon (within the next 6 mo. to 1 yr.) a workable set of virtual observatory (VO) tools, including SED and time series building, displaying and analysis tools, integrated in the existing IVOA infrastructure, should be the immediate focus of the VAO.
- High-quality 21st-century data visualization and statistical tools are needed at the VAO level.
- We endorse the VAO team proposal that tools, whenever possible, should be provided with IVOA standard interfaces, so that can be accessed e.g. as part of a SAMP workflow.
- Discussion is needed with the wider community (e.g., the LSST project) on integrating more software applications (e.g., image processing, time series analysis, multivariate classification) for use with VAO extracted data.
- The VAO should take full advantage of major public-domain software packages developed outside of astronomy to leverage astronomical science analysis.
- The most important aspect of the above is a focus on using what has already been created, within and beyond the NVO effort, to "connect" useful pieces into a working ecosystem for scientists.

**1.2 Stronger Support for New Visions and Technologies**

The VAO-SC is concerned that not enough emphasis has been given in the formulation of the VAO to the exploration and uptake of new technologies (only a fraction of FTE is allocated in the budget to this function). The VAO will need to stay abreast of the fast-changing technical landscape, to be successful in its endeavor. *We recommend stronger emphasis on this function.*

Visualization, the ability of using workflows, and connectivity – data, tools, literature - are all emerging as important areas for the VAO. While the starting concept of the VO was about distributed data and a common system to access it, the present paradigm is moving towards distributed tools accessing common data centers. This approach is being recognized by the VAO team, and we endorse it. Capability and ease-of-use improvements to the web typically now come in the form of nesting, aggregating or connecting tools. In the *Seamless Astronomy* view, today's VO should be thought of as the ever improving set of data archives, tools, interconnections, and standards that strive to make astronomical research as seamless as travel research. Software tools are becoming much more interoperable thanks to protocols for message passing, such as SAMP. The improved speed of web applications is to some extent removing platform-dependence as an obstacle to programmers and users alike. An ever-increasing diversity of tools allow researchers to carry out a particular research task, so that the important research for the future lies in figuring out how to make the tools, their interconnections, and their connections to data and literature resources useful and well-known to the astronomical community. *Providing to users existing and emerging technologies in these field, is a crucial area of VAO activity.*

**1.3 Improved Visibility in the Scientific Community**

The VAO must have an active and multi-faceted program of community involvement, ranging from a more visible presence at the AAS, to tutorials at different astronomy centers throughout the country. The successful NVO summer school should be continued and used to prepare the material for these more astronomy-based outreach efforts. Web-based tutorials, including video screen-cast demos, and online descriptions and discussions of software tools should be continued and significantly augmented. It is important in these activities to listen to the user community, especially the majority of potential users who are not VO-cognizant. We urge testing these materials in a variety of environments and changing them in response to users feedback. VAO user groups at major US astronomical facilities should be created and used as a source of input.

However, *summer schools and AAS presentations should not be seen as substitutes for producing a Web Portal that is both well documented and intuitively obvious to use*. The first interactive encounter that potential users have with the VAO will be critical; they will not come to a summer school for advanced projects if they cannot get something quickly and easily from the VAO website on their first try. The VAO website should be well tested, robust and self-documenting; no training required. The VAO-SC recommends that the VAO website be designed by a team with proven web design and communication expertise, possibly including an outside user-interface/web-usability consultant. Being an excellent working astronomer or IT expert does not provide the ability of designing an effective portal.

In order to track the use of its facilities, the VAO should urge users to acknowledge the use of VAO tools and facilities in their publications.

## 1.4 Interfaces with Existing and Upcoming Data Centers

For the Virtual Observatory to be successful, it is essential that VO standards be used by all major established (e.g., the NASA Data and Mission Centers) and soon to become operational (e.g., the NSF ALMA) data centers. The VAO needs to be visible to and collaborate with the large ground-based survey teams (LSST, PanSTARRS). Centers/archives for Exoplanet research should also be included. We recognize that the VAO-SC does not have the resources, or the scope, to impose such an uptake. However, it is important that communications at various levels are established early-on with all these major centers and projects, to foster either transition to VO standards or the adoptions of these standards in the design of new data systems.

Moreover, it is important that the VAO ensure that data centers acknowledge the use of VO standards whenever appropriate, and make their users aware of this policy, so that the VAO can get proper recognition. Reciprocally, the VAO should be careful to track, cite and properly display provenance (history of ownership and origin) when it uses tools, provides external services or accesses and repackages data that reside at other sites, centers or websites.

Acknowledgement, in online and published work, of the VAO should become analogous to acknowledgement of a traditional observing facility or funding agency.

## 1.5 Channels for Science Funding

The VAO has no clear provision in its budget for the funding of scientific projects by community members. However, the presence of 'guest investigator' programs in the large NASA observatories is one of the reasons for their great success, by engaging the community and allowing for a new generation of 'expert' astronomers to be raised. These thriving communities provide feedback to the Data Centers, stimulating more efficient operations and archives, and the development of new science-motivated analysis tools.

It is particularly important to provide these opportunities to the scientific community, especially in the initial phase of operations of the VAO, to provide an added incentive to the use of VAO facilities. The VAO-SC realizes that different approaches to science funding are followed by the two funding agencies, NASA and NSF. Nevertheless, the SC urges the VAO Director to work with the Agencies to find a way to address the issues of funding for scientific research, so that mechanisms may be set in place to issue a call for proposals for VAO research. These proposals can either be around developing special data sets, or developing special, re-usable, tools, or both.

The VAO-SC urges the Director to establish, whenever appropriate, partnerships with commercial companies active in computing technologies. For example, Microsoft would be an appropriate partner, and so would be Google, just to name two obvious candidates. These partnerships could be a mechanism for establishing centers of VAO expertise, and providing funding for VAO activities.

The "VAO Associates" program mentioned by Director Hanisch at our meeting seems to lead exactly in the right direction, and should be funded.

## 2. RECOMMENDATIONS FOR NEAR-TERM IMPLEMENTATION

The VAO-SC has identified several areas that the VAO needs to address urgently, in the first year of operations. We encourage the VAO to move very rapidly towards delivering some new, unique and useful science-driven services within the first six months of funding.

### 2.1 Science Community Education and Visibility

**2.1.1 Improve the user interface** – The NVO portal gives a limited view of the VO at large. Since developing the VO infrastructure has been and continues to be a worldwide community effort, the VAO portal should give equal access to all tools and facilities as appropriate, not just focus on the NVO-developed tools. These NVO-developed tools were considered somewhat rough, in comparison with other development from international (e.g. TopCat; SAMP) and independent (e.g., astrometry.net) groups, eliciting the comment that all the action is elsewhere. We recognize that this is an unfair comment since the NVO has contributed significantly to the VO infrastructure, but it has relevance for how the VAO presents itself to the science community. The EuroVO web page emerged in the discussion as a possible model for engaging the scientific community. We urge that the VAO:
- Consider these other VO pages when re-examining the portal. There is nothing wrong with a US "portal" guiding users to internationally developed tools.
- Work with community outreach experts in this area. *Technical expertise does not ensure communication expertise.*

**2.1.2 Provide science user support** – Although user support is part of the VAO proposal, the VAO-SC believes that a missing persona in the VAO is the *VAO astronomer*, a cadre of enthusiastic astronomers with data savvy and demonstrated knowledge of advanced analysis and statistical methods, engaging in VO science and expert of all VO tools and facilities (VAO and other). All successful major observatories/data centers have these expert astronomers. These are the people that will interface with the users and help them out; also they would be the 'external' testing team. This does not need to be a centralized function; centers of VAO expertise could be spread out geographically, to best serve their immediate local communities. These positions are analogous to the "friend of the telescope" employed by traditional observatories.

**2.2 Data-Publication Connection**

With ADS the astronomical community has a powerful literature search engine, which has been extended to include data links and to perform semantic searches. Some of these functions are however limited (or made unwontedly laborious) by the incomplete keywords and indexing of the journal papers. This problem also affects the flexibility of searches with the well-established astronomical service NED.
The VAO-SC recommends that the VAO early on:
- **Establish a liaison to the journal publication boards** (ApJ and AJ), to raise awareness of these issues and work out solutions.
- **Enhance the data publishing capability** for these journals by developing a data publishing tool, to be used during the paper submission process by authors.

**2.3 User Tools and Services**

The VAO-SC recognizes that there is infrastructure work still ongoing; however, the existing infrastructure is mature enough to allow some immediate implementations that will help buying in by the astronomy community. The areas of more immediate concern are the coordination of time and spectral domains. *We point out that whenever possible the VAO should not produce new tools, but investigate and adapt what is already available at the (national and international) data and archival centers*. These tools should be SAMP-enabled, so as to be used as part of workflows in conjunction with visualizations tools and ADS for the literature connection. We acknowledge that discussions of these areas (e.g., SED tools, tools to cast models in observation space, fitting and cross-matching tools) were presented by the VAO team to the VAO-SC, during our meeting. We endorse these plans and identify the following areas for immediate attention.

**2.3.1 SED tools** – These tools include:
- **SED builders**, tools to extract spectral/photometric data from different data sets, calibrate the data and generate the SED to be used for further analysis. These tools may include interactive display components. We note that SEDs in the VO context extend from radio to gamma-rays.

- **SED analyzers**, including classifier tools to compare SEDs to established templates; and statistical analysis and model fitting tools, to provide a measure of goodness of the fit, and allow a flexible model interface.
- **SED display** tools

**2.3.2 Time series tools** - Time series tools closely reflect the functions outlined for SED tools, although calibrations will be time-axis specific, and additional analysis tools (e.g. Fourier transforms) may be required in the analysis package. Close collaboration with existing efforts (e.g. within Pan-Starrs, LSST and the Harvard Time-Series Center) should be sought in these efforts.

**2.3.3 Image display and visualization tools** – The VAO should allow easy comparison of data acquired by instruments at widely different angular resolution. The image display facility (e.g. DS9) should offer the capability for easily visualizing data at a specified 'similar' angular resolution. Ability to degrade PSFs `at will', or to measure them on-the-fly and use the measurements to bring different images onto a similar angular scale is going to be crucial also for new (Herschel Space Telescope) and upcoming (ALMA) facilities.

**2.3.4 User-friendly cross-matching interface** – Existing user interfaces tend to be specific to a particular database (e.g. SDSS), putting the burden of learning the intricacies of all these interfaces on the user who wants to engage in multi-wavelength research. Moreover, the casual or beginner user should not be required to be expert in SQL. We urge the VAO to provide a user-friendly search interface that could cross-match several VO-published databases.

**2.4 Key VAO Science and Data Products**

The establishment of *key VAO science projects* is an area that the VAO-SC discussed as a way to get the community (and SC members) involved in the VAO as users, do science and at the same time validate the science tools. It is however an area of which the Directors will need to consider the political implications, and establish clear guidelines, addressing for example,
- How is a key project defined and who defines it?
- Once a key science project is identified, should there be a general call for proposals?
- These activities will require some funding. Are there funds available?
- Since VAO-SC members will require some support to engage in these projects, how do we ensure that the VAO-SC is not seen as a group of insiders reaping the benefits?

The discussion was centered on the use of upcoming public HST data with wide coverage at other wavelengths (e.g., M31 survey, SMC survey, lensing clusters, deep survey fields), and adding multi-wavelength cross-matching, followed e.g. by SED or time analysis. Some SC members expressed interest in getting involved in this analysis and produce papers for publications, as well as data products/catalogs for public release.

An **example** of this 'key-project-style' science (by Barry Madore) is to produce a cross-matched catalog of stars in the SMC, constructed by combining
- Zaritsky's ground-based UBVRI imaging catalog with
- 2MASS JHK data for point sources in the same region,
- Spitzer/SAGE SMC 3.6, 4.5, 5.8 and 8.0 micron imaging survey and
- OGLE-III time-series V & I-band survey comprising over 400 million stars in the SMC observed typically 700 times in the I-band and 50-70 times in the V-band.

In this way the project touches on ground-based data and space-based data, merges European and American datasets, and even involves the time-domain and the possibility of a future interaction with theoretical modeling. From this master catalog we will then be able to extract the phased multi-wavelength light curves of over 4,600 known Cepheids. And using reddening-corrected, mean-magnitude Period-Luminosity relations we will have both the dense spatial sampling and the necessarily very precise individual distances needed to derive the three-dimensional structure of the SMC. The resulting study will act as an important constraint on interaction models involving the production of the Magellanic Stream and the mutual interaction and structural evolution of the Large and Small Magellanic Clouds. The merged catalog could be served by IRSA, while the extragalactic Cepheids could be assimilated into NED and perhaps SIMBAD.

## 3. SCIENCE INPUT INTO VAO PLANNING

The VAO-SC looks forward to the draft of the VAO Program Plan, identifying schedules and milestones for the delivery of science capabilities to the community, and is ready to work with the management team to provide feedback and further science input. In addition, the VAO-SC urges the VAO management to formulate longer-term plans. We recommend that a 5-yr plan be generated, and we are willing to review it and provide science input to the priorities. We urge the VAO to include Data Mining in the planning for future development. *The VAO-SC is ready to provide topical expertise (e.g. statistics, visualization, data mining), and assist the planning and technical team in achieving the recommended goals.*

### 3.1 Near-term Delivery Planning

We urge the VAO to include in the VAO Program Plan plans for the delivery of the science tools and capabilities identified in this VAO-SC report as urgent (Section 2.). We urge that this plan be provided to the VAO-SC within the first 2 mo. of VAO operations for timely science feedback to the program team. We suggest that this plan identify clearly
- Science capabilities to be expected at the end of the first 6 mo. of operations,
- Additional science capabilities to be expected at the end of the first year of VAO operations,

- A revised personnel plan, showing differences between what was envisioned in the original VAO proposal in terms of staffing needs, and the situation now.

**3.2 The 5-Year Plan**

The VAO-SC strongly recommends that the VAO prepare and submit to the SC – in the same time scale - for review a longer-term, 5-year plan, outlining the major milestones envisaged for the project, with expected delivery dates to the users. The importance of this plan is to ensure that the VAO has a strategic vision, and that this long-term vision receives science vetting. We understand that this plan, especially as it covers the more distant future, will in all likelihood be modified, in response to the changing scientific and technical landscape.

**3.3 Report on evaluation of new approaches**

We urge the VAO prepare a report for the VAO-SC that evaluates approaches to improved and up-to-date (1) data visualization; (2) statistical analysis; and (3) data mining software, based on both intra-VAO software and integration of existing packages for VAO use. The capabilities should be on the level of the latest public domain capabilities available for medical imaging, remote sensing, engineering, and statistics. The goal would be to produce or integrate capabilities that go considerably beyond those provided by commonly used software today (e.g. IDL). Plans for community training in new tools, such as incorporation into the NVO Summer School, should be outlined.

In particular, Data Mining is an emerging discipline, which will foster unanticipated discoveries, by comparing the large multi-wavelength data sets that the VO is making accessible. We urge the VAO to include Data Mining in the planning for future development. We would like to see a plan that will start now on the path towards achieving a full workable set of data mining capabilities five years from now, including step-wise yearly deliveries of tools, and programs for tutorials, workshops, and community involvement.

The VAO-SC volunteers to provide their expertise to the VAO to achieve these recommended science goals.